\documentclass[letterpaper]{article} 
\usepackage{aaai25}  
\usepackage{times}  
\usepackage{helvet}  
\usepackage{courier}  
\usepackage[hyphens]{url}  
\usepackage{graphicx} 
\usepackage{natbib}  
\usepackage{caption} 
\usepackage{algorithm}
\usepackage{algorithmic}
\usepackage{comment}
\usepackage{newfloat}
\usepackage{listings}
\usepackage{multirow}
\usepackage{array}
\usepackage{makecell}
\usepackage{booktabs}
\usepackage{enumitem}
\usepackage{times}
\usepackage{latexsym}

\usepackage[dvipsnames]{xcolor}         
\usepackage{tcolorbox}

\usepackage{color, colortbl}
\definecolor{deepred}{rgb}{0.631,0.102,0.102}
\definecolor{gyellow}{HTML}{F4B400}
\definecolor{mildyellow}{HTML}{FFF2CC}

\newtcolorbox{GenerateTextualFinancialStatements}{
  colback=gyellow!10,
  colframe=gyellow!30!black,
  fonttitle=\bfseries,
  title=Templates to Generate Textual Financial Statements,
  sharp corners,
}

\newtcolorbox{GenerateSyntheticTransactionData1}{
  colback=gyellow!10,
  colframe=gyellow!30!black,
  fonttitle=\bfseries,
  title=Templates to Generate Synthetic Transaction Data,
  sharp corners,
}

\newtcolorbox{GenerateSyntheticTransactionData2}{
  colback=gyellow!10,
  colframe=gyellow!30!black,
  fonttitle=\bfseries,
  title=Templates to Generate Synthetic Transaction Data,
  sharp corners,
}

\newtcolorbox{TemplatesforErrorInjectioninFinancialStatements1}{
  colback=gyellow!10,
  colframe=gyellow!30!black,
  fonttitle=\bfseries,
  title=Templates for Error Injection in Financial Statements,
  sharp corners,
}

\newtcolorbox{TemplatesforErrorInjectioninFinancialStatements2}{
  colback=gyellow!10,
  colframe=gyellow!30!black,
  fonttitle=\bfseries,
  title=Templates for Error Injection in Financial Statements,
  sharp corners,
}

\newtcolorbox{TemplatesforErrorInjectioninFinancialStatements3}{
  colback=gyellow!10,
  colframe=gyellow!30!black,
  fonttitle=\bfseries,
  title=Templates for Error Injection in Financial Statements,
  sharp corners,
}

\newtcolorbox{TemplatesforPromptingSotaLLMstoServeasAuditors1}{
  colback=gyellow!10,
  colframe=gyellow!30!black,
  fonttitle=\bfseries,
  title=Templates for Prompting Sota LLMs to Serve as Auditors,
  sharp corners,
}

\newtcolorbox{TemplatesforPromptingSotaLLMstoServeasAuditors2}{
  colback=gyellow!10,
  colframe=gyellow!30!black,
  fonttitle=\bfseries,
  title=Templates for Prompting Sota LLMs to Serve as Auditors,
  sharp corners,
}

\usepackage[T1]{fontenc}

\usepackage[utf8]{inputenc}

\usepackage{microtype}

\usepackage{inconsolata}

\usepackage{graphicx}
\usepackage{newfloat}
\usepackage{listings}
\DeclareCaptionStyle{ruled}{labelfont=normalfont,labelsep=colon,strut=off} 
\floatstyle{ruled}
\newfloat{listing}{tb}{lst}{}
\floatname{listing}{Listing}
\title{Automating Financial Statement Audits with Large Language Models}
\author{
    Rushi Wang\textsuperscript{\rm 1}, Jiateng Liu\textsuperscript{\rm 1}, Weijie Zhao\textsuperscript{\rm 2}, Shenglan Li\textsuperscript{\rm 2},
    Denghui Zhang\textsuperscript{\rm 2}\\
}
\vspace{0.5em}
\affiliations{
    \textsuperscript{\rm 1}University of Illinois Urbana-Champaign, 
    \textsuperscript{\rm 2}Stevens Institute of Technology\\
    \vspace{0.5em}
    \{rushiw2, jiateng5\}@illinois.edu, dzhang42@syevens.edu.
}

\usepackage{xcolor}

\usepackage{xcolor}

\usepackage{bibentry}

\begin{document}

\maketitle

\begin{abstract}
Financial statement auditing is essential for stakeholders to understand a company's financial health, yet current manual processes are inefficient and error-prone. Even with extensive verification procedures, auditors frequently miss errors, leading to inaccurate financial statements that fail to meet stakeholder expectations for transparency and reliability.
To this end, we harness large language models (LLMs) to automate financial statement auditing and rigorously assess their capabilities, providing insights on their performance boundaries in the scenario of automated auditing. Our work introduces a comprehensive benchmark using a curated dataset combining real-world financial tables with synthesized transaction data. 
In the benchmark, we developed a rigorous five-stage evaluation framework to assess LLMs' auditing capabilities. 
The benchmark also challenges models to map specific financial statement errors to corresponding violations of accounting standards, simulating real-world auditing scenarios through test cases.
Our testing reveals that current state-of-the-art LLMs successfully identify financial statement errors when given historical transaction data. However, these models demonstrate significant limitations in explaining detected errors and citing relevant accounting standards. Furthermore, LLMs struggle to execute complete audits and make necessary financial statement revisions.
These findings highlight a critical gap in LLMs' domain-specific accounting knowledge. Future research must focus on enhancing LLMs' understanding of auditing principles and procedures. Our benchmark and evaluation framework establish a foundation for developing more effective automated auditing tools that will substantially improve the accuracy and efficiency of real-world financial statement auditing. 

\end{abstract}
\section{Introduction}

\begin{figure*}[h]
    \centering    \includegraphics[width=1.0\textwidth]{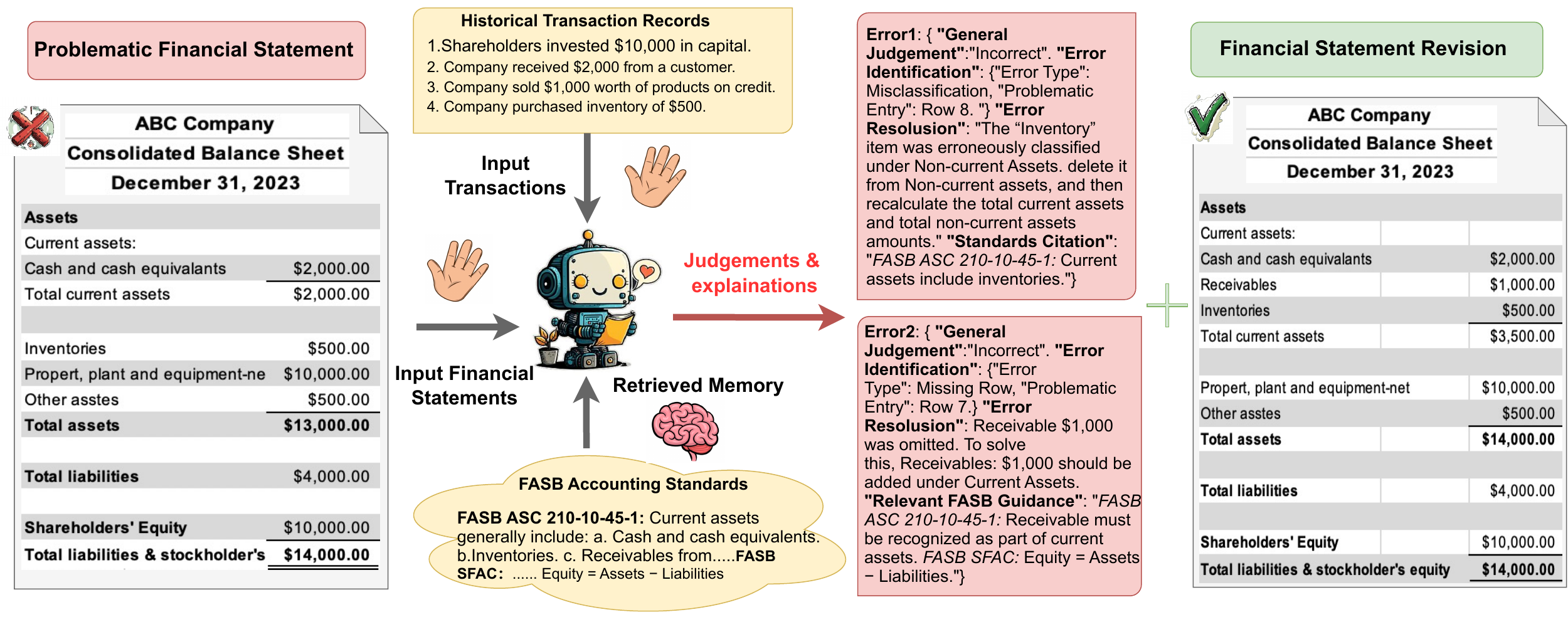}
    \caption{Our Automated Financial Statement Auditing pipeline: An LLM-based agent is provided with (a) The history transaction data of a company. (b) Financial statements prepared by accountants which may potentially contain errors. The LLM-based agent needs to (1) \textbf{Retrieve} the accounting standards provided by \textbf{FASB} (Financial Accounting Standards Board). (2) \textbf{Compare} the transaction data with the financial statements and check whether they align with accounting standards. (3) \textbf{Generate} a judgment of the financial statement. \textbf{Explain} the errors and concerns to the current financial statements, and further suggest a revision. (4) \textbf{Revise} the problematic financial statements and ensure compliance with accounting standards.}
    \label{fig:main}
\end{figure*}


In today's interconnected global financial market, auditing financial statements is essential to ensure that stakeholders have a transparent understanding and accurate information on a company's financial health~\cite{popovic2015importance}. However, the current workflow for auditing financial statements is neither efficient nor accurate. Despite the transformative impact of information technologies like auditing software~\cite{debreceny2005employing} and big data analytics~\cite{cao2015big,stewart2015data} on the auditing industry, most audit work remains heavily reliant on human labor~\cite{byrnes2018evolution} and the process is characterized by extensive labor, inefficiency, and less-than-optimal accuracy~\cite{cho2022auditors,li2023government}. 
Meanwhile, auditors in the current workflow may only focus their efforts on areas that are more likely to have material misstatements or fraud, potentially neglecting less granular or lower-priority details,  which may result in audited financial statements inaccuracy while failing to meet user expectations~\cite{toumeh2018expectations}. Furthermore, the exponential growth in data volume and complexity intensifies these challenges~\cite{sun2024study}, underscoring the urgent need for intelligent agents to enhance both the accuracy and efficiency of financial statement audits.

Previous works for simplifying the auditing processes mostly focused on better sampling techniques~\cite{elder2013audit,nigrini2017audit} or efficient data extraction approaches~\cite{sirikulvadhana2002data,vinatoru2016using}, which address only specific aspects of the auditing workflow. However, recent advancements centered on Large Language Models (LLMs) have pushed the boundaries of automation in auditing financial statements. Their strong prior knowledge~\cite{hu2023survey} and capability in reasoning~\cite{wei2022chain} enable improvements in tasks such as identifying inconsistencies in financial narratives through text matching~\cite{hillebrand2023improving} and ensuring adherence to complex regulatory frameworks through automated compliance verification~\cite{berger2023towards}. \textbf{Nevertheless, these approaches still fail to address low-level auditing challenges, such as the labor-intensive cross-verification of transaction data against financial disclosures. Additionally, it is crucial that the auditing model understands and strictly follows accounting standards. The model should also provide justifications to ensure the trustworthiness of the output financial statements.} Our work is to tackle this foundational issue, offering a solution that systematically evaluates the LLM's abilities in auditing financial statements, particularly focusing on tasks such as identifying and cross-verifying transaction data with financial disclosures to address critical low-level auditing challenges.



This paper explores enhancing LLMs for automated financial statement auditing. As illustrated in Figure~\ref{fig:main}, the model analyzes historical transaction data and a generated financial table, which may be accurate or contain errors. Its objective is to verify the table's accuracy, assess alignment with transaction data, identify discrepancies, and provide detailed justifications. Recognizing the critical role of financial auditing, we propose a systematic five-stage diagnostic framework to evaluate the model's performance.

\begin{enumerate}
    \item \textbf{General Judgment}: Provide an overall assessment of the correctness of the financial statements, framed as a binary classification task (correct/incorrect).
    \item \textbf{Error Identification}: Identify the error types in the statements and the specific erroneous entries within the financial table.
    \item \textbf{Error Resolution}: Analyze the identified errors and propose amendments to address the identified issues using natural language explanations accordingly.
    \item \textbf{Standards Citation}: Generate references to relevant accounting standards as evidence to substantiate findings.
    \item \textbf{Financial Statement Revision}: Apply direct modifications to the table and generate a corrected version of the financial statements.
\end{enumerate}


Based on our framework, we introduce the first benchmark for automatic financial statement auditing using real financial statements from S\&P 500 companies globally. We carefully curate this data manually and produce high-quality synthetic historical transaction data. Our data generation process aims to maintain a balanced mix of correct and erroneous tables and to diversify the types of discrepancies between transaction data and financial statements. This approach enhances the realism and complexity of our benchmark, making it a robust tool for evaluating the efficacy of automated auditing systems.

With the created benchmark together with our evaluation framework, we conduct comprehensive experiments for evaluating the capabilities of State-of-The-Art 
LLMs on financial statement auditing. We first verified that current LLMs are sufficient for identifying errors in financial statements when provided with historical transaction data. However, they still need improvement in terms of generating a clear explanation of the error and providing accounting standards evidence. We draw valuable qualitative analyses and provide insights into how we can potentially improve the abilities of LLMs on this specific task, which should pave the way for further research in the related area and push it towards real-world applications.




Overall, our contributions are: 
\begin{itemize}
    \item We introduce a novel task for the automatic auditing of financial statements, highlighting critical low-level auditing challenges and laying the foundation for enhanced accuracy and efficiency in real-world auditing practices.
    \item We develop the first high-quality benchmark for the automatic auditing of financial statements, offering a valuable resource to drive future research in this domain.
    \item We conduct a systematic evaluation of LLMs’ capabilities in automatic financial statement auditing, providing detailed qualitative analyses to inspire future advancements.
\end{itemize}

\section{Automatic Auditing Benchmark}

We release the first benchmark for evaluating LLMs' ability of Automatic Auditing of Financial Statements, aiming to advance research and applications in this domain. For each data piece, input data contains a pair of the generated financial statements and the company's historical transaction data. The financial statements are labeled as either correct or incorrect based on their alignment with the transaction data. We also provide the ground truth five-stage evaluation results as part of the dataset. Additionally, we provide a curated knowledge base of accounting standards, serving as the agent's external memory. We have professional accounting practitioners manually summarize the accounting standards as a comparison to the original standards. While the original standards offer comprehensive details, the summaries are easier for model to understand. 

\subsection{Data Creation} 

\paragraph{Statements.}
For financial statements, we manually selected financial statements from consolidated income statement, consolidated balance sheet, consolidated statements of shareholders’ equity, and consolidated statements of cash flows from randomly selected companies in the S\&P 500.  The table data are captured as screenshots and then transformed into structured text using predefined rules. Each table begins with a [Tab] token and a descriptive title, followed by year-specific data introduced by [Time] tokens. Rows are indexed as ([row n]), with columns separated by vertical bars \(|\), and concluded with [SEP] tokens. We execute this process with GPT-4-vision while a notable number of errors occurred during the transformation. As a result, we manually check and modify a small set of tables and ensure the accuracy of the structured representations. An example of a transformed table is shown in the Appendix. 

\paragraph{Transactions.} Historical transaction data from real-world companies are hardly applicable, as the majority of such data are confidential. As a result, we asked accounting major students to write few-shot examples to prompt the GPT-4 models to generate diversified synthetic transaction data and then followed by a round of human verification. While we ensured the generated data is of high quality, the amount and truthfulness of the data were insufficient. These limitations are further discussed in the limitations section. 

\paragraph{Errors and Explanations.}
To create erroneous tables, we first define the error types for financial statements. Comparing a wrong table with a right table, there are the following types of errors: \textbf{(1) Missing Row}: A row containing a specific account with values is missing. \textbf{(2) Numerical Error}: The value of a specific account is incorrectly computed. \textbf{(3) Redundant Row}: A redundant row with a logical account name and amount that fits the respective category \textbf{(4) Misclassification}: A wrong row from one category is written to another where it logically does not belong. (e.g., "Accounts Payable" from "Liabilities" is moved to "Assets"). Based on the error types, we ask GPT-4 to insert these errors into the original table. We take single errors and multiple errors into consideration. For multiple errors, we ask the GPT models to insert several errors from different categories into the original table and provide information for these errors sequentially. 
We manually verify the generated errors and make sure that the concrete errors are diverse. Along with the erroneous tables, we generate five-stage evaluation ground truth labels simultaneously. Since the GPT models are the generator of these errors, it is easy to ensure that the ground truth labels are aligned and accurate.
Prompts for generating synthetic data are all provided in the Appendix.

\subsection{Dataset Statistics}

We manually collected 124 images form from the S\&P500 Companies. Since each image may contain financial statements from different years, we manually curate them into 371 financial statements expressed in textual format. Among them, there are 101  consolidated balance sheets, 140 consolidated statements of income, and 130 consolidated statements of cash flows. We did not include the consolidated statements of shareholders' equity since they are hard to parse into text. We first insert only one type of error that is mentioned above in each of the tables, resulting in 1484 wrong tables. We also insert random multiple errors into the tables, resulting in 371 wrong tables which are more difficult for LLMs.









\section{Experiments}

\subsection{Experimental Settings} 

We utilize state-of-the-art large language models GPT-3.5-turbo and GPT-4 to assist in auditing financial statements. Evaluations are conducted independently across financial statements that are correct, contain a single error, and contain multiple errors. Our experiments do not differentiate between various types of financial sheets.

\subsection{Evaluation Metrics}
We used the following metrics to evaluate the five stage's performance, besides, we also added an extra overall evaluation which synthesizes scores from all stages to generally judge the model performance.

\begin{itemize}
    \item \textbf{Evaluate General Judgment}: The binary classification task is evaluated with an Exact Match Score (EM Score).
    \item \textbf{Evaluate Error Identification}: The type of the error and the entry position of the problematic table are independently evaluated with EM Score.
    \item \textbf{Evaluate Error Resolution}: As a generation-based task, error resolution is evaluated using BertScore, which assesses the semantic similarity between the generated results and the ground truth.
    \item \textbf{Evaluate Standards Citation}: For this stage, an external retriever extracts relevant accounting standards from our standards database. The evaluation compares the retrieved standard's index with the ground truth using the Top-K retrieval-based EM Score.
    \item \textbf{Evaluate Financial Statement Revision}: The modifications to table entries are compared to the ground truth using BLEU, which focuses on capturing low-level accuracy in the generation process.
    \item \textbf{Evaluate the Overall Auditing Process}: The complete auditing process is assessed with Success Rate (SR), which measures the model's ability to perform well in every auditing stage. We apply strict requirements to the output. We require all the EM-Score to be 1, the BertScore to be over 0.85, and the BLEU score to be over 0.99 to obtain a complete successful auditing.
\end{itemize}

\begin{table}[ht]
\centering
\caption{The table shows the state-of-the-art LLMs: GPT-3.5 Turbo and GPT-4 performance on identifying correct financial statements.}
\label{tab:tab3}
\resizebox{0.45\textwidth}{!}
{
    \begin{tabular}{cccc}
    \toprule
    \textbf{Evaluation $\rightarrow$} & \multicolumn{1}{c}{\textbf{General Judgment}} & \textbf{Evaluation $\rightarrow$} & \multicolumn{1}{c}{\textbf{General Judgment}} \\
    \textbf{Model $\downarrow$} & \makecell{Avg. EM Score} & \textbf{Model $\downarrow$} & \makecell{Avg. EM Score} \\
    \midrule
    GPT-3.5-Turbo & 1.000 & GPT-4 &  1.000 \\
    \bottomrule
    \end{tabular}
}
\end{table}

\subsection{LLMs Effectively Avoid Misclassification of Error-Free Financial Statements} Above all, we evaluate whether state-of-the-art LLMs can accurately recognize error-free financial statements. The experimental results in Table~\ref{tab:tab1} demonstrate that LLMs effectively avoid misclassifying accurate financial statement tables as erroneous. Despite the limitations in the range of our test data, we conclude that it is safe to use LLMs as an initial step for auditing tables, as they do not introduce additional labor due to misclassification.

\begin{table*}[ht]
\centering
\caption{ This table shows state-of-the-art LLMs' auditing performance on financial statements with a \textbf{single error}. \textbf{Up:} Model performance on General Judgment, Error Identification, and Error Resolution. \textbf{Down:} Model performance on Standards Citation, Table Revision, and Overall Evaluation. We report the average performance across 150 random samples in our dataset.}
\resizebox{1.0\textwidth}{!}
{
    \begin{tabular}{lccccc}
    \toprule
    \textbf{Evaluation $\rightarrow$} & \multicolumn{1}{c}{\textbf{General Judgment}} & \multicolumn{1}{c}{\textbf{Error Type Identification}} & \multicolumn{1}{c}{\textbf{Error Entry Identification}} & \multicolumn{1}{c}{\textbf{Error Resolution}} \\
    \textbf{Model $\downarrow$} & \makecell{Avg. EM Score} & \makecell{Avg. EM Score} &  \makecell{Avg. EM Score} & \makecell{Avg. BertScore} &   \\
    \cmidrule{2-3} \cmidrule{4-6}
    GPT-3.5-Turbo & 1.000 & 0.764 & 0.418  & 0.869  \\
    GPT-4 & 1.000 & 0.899 &  0.737 & 0.878 \\
    \midrule
    \textbf{Evaluation $\rightarrow$} & \multicolumn{1}{c}{\textbf{Standards Citation(Top-1)}} & \multicolumn{1}{c}{\textbf{Standards Citation (Top-5)}} & \multicolumn{1}{c}{\textbf{Table Revision}} & \multicolumn{1}{c}{\textbf{Overall Evaluation}} \\
    \textbf{Model $\downarrow$} & \makecell{Avg. EM Score} &  \makecell{Avg. EM Score}  & \makecell{Avg. BLEU} &  \makecell{Avg. Success Rate} &   \\
    \midrule
    GPT-3.5-Turbo & 0.137 & 0.360 & 0.707 & 0.025  \\
    GPT-4 & 0.262 & 0.515  & 0.783 & 0.041\\
    \bottomrule
    \end{tabular}
}

\label{tab:tab1}
\end{table*}

\subsection{LLMs Fall Short of Meeting Real-World Financial Auditing Requirements} We conducted comprehensive evaluations of state-of-the-art LLMs based on our evaluation metrics. The experimental results for auditing financial statements containing a single error are presented in Table~\ref{tab:tab2}, while the results for auditing financial statements with multiple errors are shown in Table~\ref{tab:tab3}. The findings reveal that although both GPT-3.5-turbo and GPT-4 excel at identifying misalignments between transaction data and generated financial statements, their performance significantly declines for more complex tasks, such as identifying specific error types. For instance, GPT-4 achieves only about 50\% accuracy in this task. Moreover, the accumulation of errors in 'Error Type Identification' and 'Error Entry Identification' makes it extremely challenging for the models to revise tables accurately. This limitation becomes particularly evident in real-world audit scenarios, where auditors are required not only to detect errors in financial statements but also to provide corrected versions. We observed a further decline in performance when the target financial statements contained multiple errors. We hypothesize that this performance drop would become even more pronounced in real-world scenarios, where errors are both prevalent and irregular.
Overall, the results indicate that current state-of-the-art LLMs are insufficient to serve as reliable auditors within the auditing workflow.


\begin{table*}[ht]
\centering
\caption{ This table shows state-of-the-art LLMs' auditing performance on financial statements with \textbf{multiple errors}. \textbf{Up:} Model performance on General Judgment, Error Identification, and Error Resolution. \textbf{Down:} Model performance on Standards Citation, Table Revision, and Overall Evaluation. We report the average performance across 150 random samples in our dataset.}
\label{tab:tab2}
\resizebox{1.0\textwidth}{!}
{
    \begin{tabular}{lccccc}
    \toprule
    \textbf{Evaluation $\rightarrow$} & \multicolumn{1}{c}{\textbf{General Judgment}} & \multicolumn{1}{c}{\textbf{Error Type Identification}} & \multicolumn{1}{c}{\textbf{Error Entry Identification}} & \multicolumn{1}{c}{\textbf{Error Resolution}} \\
    \textbf{Model $\downarrow$} & \makecell{Avg. EM Score} & \makecell{Avg. EM Score} &  \makecell{Avg. EM Score} & \makecell{Avg. BertScore} &   \\
    \cmidrule{2-3} \cmidrule{4-6}
    GPT-3.5-Turbo & 1.000 & 0.482 & 0.349 & 0.637 \\
    GPT-4 & 1.000 & 0.752 & 0.587 & 0.792 \\
    \midrule
    \textbf{Evaluation $\rightarrow$} & \multicolumn{1}{c}{\textbf{Standards Citation(Top-1)}} & \multicolumn{1}{c}{\textbf{Standards Citation (Top-5)}} & \multicolumn{1}{c}{\textbf{Table Revision}} & \multicolumn{1}{c}{\textbf{Overall Evaluation}} \\
    \textbf{Model $\downarrow$} & \makecell{Avg. EM Score} &  \makecell{Avg. EM Score}  & \makecell{Avg. BLEU} &  \makecell{Avg. Success Rate} &   \\
    \midrule
    GPT-3.5-Turbo & 0.086 & 0.265 & 0.680  & 0.012  \\
    GPT-4 & 0.193 & 0.396  & 0.742 & 0.030\\
    \bottomrule
    \end{tabular}
}

\end{table*}


\subsection{Qualitative Analysis and Further Discussions}  To explain the observed poor performance, we identify two primary challenges hindering the application of LLMs in automatic financial statement auditing: (1) \textbf{Lack of Domain-Specific Knowledge:} Despite their advancements in general and prior knowledge, state-of-the-art LLMs lack specialized expertise in accounting standards, principles, and frameworks. Accounting and finance are highly regulated fields that demand precise interpretations of complex concepts and strict compliance with international standards. Without this domain-specific grounding, the models struggle to accurately address real-world problems, especially when dealing with nuanced or context-dependent scenarios. Our experiments reveal that it is particularly challenging for LLMs to generate accurate citations of financial standards, even within a limited scope. This underscores the persistent gap in domain-specific knowledge, even in highly advanced and otherwise knowledgeable models. (2) \textbf{Challenges in Joint Reasoning Across Tables and Text:} The models face significant difficulties in performing complex reasoning that involves synthesizing information from both tabular data and unstructured text. For financial auditing, this capability is critical, as it requires understanding interrelations between numerical data, textual explanations, and contextual cues across multiple documents. Current LLMs often treat these modalities independently, leading to fragmented reasoning and incorrect conclusions.
In our experiments, we observe that models struggle to accurately locate errors in tables and correct them, even when provided with a predicted error cause, highlighting the persistence of this issue. 

\paragraph{Suggestions for Future Research:} To address the challenges identified, future work should focus on two key areas: (1) integrating domain-specific knowledge in accounting and finance by fine-tuning LLMs on datasets enriched with accounting standards, audit procedures, and financial reports, and employing techniques like retrieval-augmented generation (RAG) or knowledge graphs for authoritative knowledge access; and (2) enhancing hybrid reasoning across tables and text through multi-modal architectures and pretraining strategies tailored for tabular data, enabling better alignment and reasoning across these modalities. Advancing these areas will improve LLMs' precision, reliability, and applicability in financial auditing and other specialized fields like healthcare and law.

\section{Conclusion}

Auditing financial statements is essential for financial transparency, yet current methods are often inefficient and error-prone. This work leverages the reasoning capabilities of large language models (LLMs) to automate and enhance the auditing process. We develop a benchmark dataset combining real-world financial tables with synthesized transaction data and systematically injected errors, showcasing LLMs’ ability to detect discrepancies in alignment with accounting standards. A five-stage evaluation framework further assesses their performance, offering insights into integrating LLMs into auditing tasks.

Our experiments show that while LLMs effectively identify errors in financial statements when paired with historical data, they struggle to generate comprehensive explanations and align findings with accounting standards. Addressing these challenges will require enhancing LLMs’ reasoning abilities, integrating domain-specific pretraining, and refining their capacity to produce evidence tied to structured accounting standards. Expanding the dataset to include diverse error types and complex scenarios is also crucial. These advancements have the potential to transform financial auditing, delivering more accurate, efficient, and scalable solutions to meet public expectations.

\newpage
\section{Limitation}

We reflect on the limitations of our work below:
\begin{enumerate}
    \item While our proposed task setting is similar to the real-world auditing scenario, the transaction data provided in our benchmark is simplified and lacks scalability. In real-world applications, auditors are faced with large amounts of transaction data, and it's unlikely we can feed this large corpus directly into the context window of current large language models. This limitation is quite prevalent in data-sparse scenarios~\cite{liu2024propainsight} However, when combined with sampling-based auditing techniques, we believe that the proposed framework can still enhance the efficiency and reliability of the auditing process.
    \item Another limitation of our study lies in its narrow focus on detecting errors in the primary financial statements using Large Language Models. In practice, auditors are required to assess a much broader range of reports, such as budget variance analyses, and segment disclosures, among others. These tasks involve additional layers of complexity, including inter-company adjustments, managerial estimates, and compliance with specialized reporting standards. Future research should investigate the capability of LLMs to handle these more diverse and complex auditing challenges to ensure practical applicability across a wider spectrum of audit responsibilities.
    \item The evaluated LLMs on a sample of 300 financial statements instead of the full dataset, may introduce discrepancies compared to results on the complete dataset. While sampling reflects common auditing practices, it may not fully capture the model’s overall performance. We neither incorporated more advanced prompting techniques like a multiple-interaction based~\cite{wang2023mint} or code-based~\cite{yang2024if} agent setting to complete the task. It is worth exploring whether there prompting frameworks will bring extra benefit to the task. Also, it is possible that a LLM updated with domain-specific knowledge would be helpful. We would further consider updating or continuing pretraining LLMs with unstructured text~\cite{liu-etal-2024-evedit,deng2024unke} to enhance its domain knowledge in accounting and finance.
\end{enumerate}

\bibliography{aaai25}

\onecolumn
\section{Appendix}
\label{appendix}

\subsection{Prompting Templates}

In this section, we provide all the templates we used to prompt GPT models to generate synthetic data. We also provide templates that we used to simulate LLMs as financial statement auditors.

\begin{GenerateTextualFinancialStatements}
    We use the following template to transform cropped financial statement images into textual table descriptions: 

\begin{quote}
You are a helpful assistant who can help transform table inputs into formmated text. Input a image which shows a table, you need to transform it into text obeying the following rules: 
(1) Start the table with [Tab], together with the title of the table, you can also add subtitle if necessary, summarize the table if the name is not provided. It should be one of Balance Sheet, Income Statement, Cash Flow Statement or Statement of Changes in Equity of the company.
(2) The table contains data from multiple years, translate the table year by year, seperate the information of each year with a   token. This means you should first focus on all the rows for a certain year (the year is usually in column) before looking at and transforming the next year. Use a [Time] together with the real time shown on table to show the specific time of the data.
(3) For each row without actual data, input a row with the title, for example: just put 'oprating activities' in a row and then [SEP]
(4) For data of each year, start each row with [row n], where n is the index of the row of the table. Then you sequentially output the content of each line, you should seperate columns with a '|' and use a [SEP] token at the end of each line. 

For example, your output can be something like: 

[Tab] Consolidated Balance Sheets from Abbive Inc.

[Time]: September 30, 2023 [SEP]
[row 0]: Net sales [SEP] 
[row 1]: Products | \$298,085 [SEP] 
[row 2]: Services | \$85,200 [SEP] 
[row 3]: Total net sales | \$383,285 [SEP] 

[Time]: September 30, 2024 [SEP]
[row 0]: Net sales [SEP] 
[row 1]: Products | \$236,085 [SEP] 
[row 2]: Services | \$89,100 [SEP] 
[row 3]: Total net sales | \$386,314 [SEP] 

[Time]: September 30, 2025 [SEP]
[row 0]: Net sales [SEP] 
[row 1]: Products | \$238,085 [SEP] 
[row 2]: Services | \$92,100 [SEP] 
[row 3]: Total net sales | \$389,235 [SEP] 

Now here is a table from the {} of the company {}, remember to directly generate your output. Now let's begin!
\end{quote}
\end{GenerateTextualFinancialStatements}

\newpage
\begin{GenerateSyntheticTransactionData1}
    Based on the textual financial statements obtained from the previous step, we simulate synthetic transactions to align with the financial entries. The following template is used to generate corresponding transaction data:

    \begin{quote}
    You are a financial data expert and a helpful assistant in generating synthetic transaction data from consolidated financial statements. Based on the input data from the following types of financial statements—Consolidated Income Statement, Consolidated Balance Sheet, Consolidated Statement of Cash Flow, and Consolidated Statement of Equity, you need to Generate Realistic Transactions that lead to the numbers in the table. The transaction data that you generate should follow the following rules: 
    (1) Align logically with the type of financial statement (e.g., revenue and expense transactions for income statements, asset and liability transactions for balance sheets, etc.)
    (2) Incorporate relatively plausible amounts and reflect realistic business activities for the specific company.
    (3) Generate event descriptions for each row of the financial statement, there should be several different events which contributes to the final number.
    (4) Generate transaction data row by row, and make sure that the sum of the generated scattered transactions is equal to the sum of each item on the financial statements.

    You need to generate the output in the following format (note that you must repeat the original input as part of your output):
    Input: Here is a [specific financial statements] from the [specific company]
    [The input of original financial table]

    Output: 
    Transactions: [List of realistic transactions generated based on the input]

    A concrete example is given here for your reference: 

Input: Here is a table from Meta Platforms

[Tab] Consolidated Balance Sheets from Meta Platforms, Inc.  [Time]: December 31, 2023 [SEP] [row 0]: Assets [SEP]  [row 1]: Cash and cash equivalents | \$41,862 [SEP]  [row 2]: Marketable securities | \$23,541 [SEP]  [row 3]: Accounts receivable, net | \$16,169 [SEP]  [row 4]: Prepaid expenses and other current assets | \$3,793 [SEP]  [row 5]: Total current assets | \$85,365 [SEP]  [row 6]: Non-marketable equity securities | \$6,141 [SEP]  [row 7]: Property and equipment, net | \$96,587 [SEP]  [row 8]: Operating lease right-of-use assets | \$13,294 [SEP]  [row 9]: Intangible assets, net | \$788 [SEP]  [row 10]: Goodwill | \$20,654 [SEP]  [row 11]: Other assets | \$6,794 [SEP]  [row 12]: Total assets | \$229,623 [SEP]   [Time]: December 31, 2022 [SEP] [row 0]: Assets [SEP]  [row 1]: Cash and cash equivalents | \$14,681 [SEP]  [row 2]: Marketable securities | \$26,057 [SEP]  [row 3]: Accounts receivable, net | \$13,466 [SEP]  [row 4]: Prepaid expenses and other current assets | \$5,345 [SEP]  [row 5]: Total current assets | \$59,549 [SEP]  [row 6]: Non-marketable equity securities | \$6,201 [SEP]  [row 7]: Property and equipment, net | \$79,518 [SEP]  [row 8]: Operating lease right-of-use assets | \$12,673 [SEP]  [row 9]: Intangible assets, net | \$897 [SEP]  [row 10]: Goodwill | \$20,306 [SEP]  [row 11]: Other assets | \$6,583 [SEP]  [row 12]: Total assets | \$185,727 [SEP]   [Time]: December 31, 2023 [SEP] [row 0]: Liabilities and stockholders' equity [SEP]  [row 1]: Accounts payable | \$4,849 [SEP]  [row 2]: Partners payable | \$863 [SEP]  [row 3]: Operating lease liabilities, current | \$1,623 [SEP]  [row 4]: Accrued expenses and other current liabilities | \$24,625 [SEP]  [row 5]: Total current liabilities | \$31,960 [SEP]  [row 6]: Operating lease liabilities, non-current | \$17,226 [SEP]  [row 7]: Long-term debt | \$18,385 [SEP]  [row 8]: Other liabilities | \$8,884 [SEP]  [row 9]: Total liabilities | \$76,455 [SEP]   [Time]: December 31, 2022 [SEP] [row 0]: Liabilities and stockholders' equity [SEP]  [row 1]: Accounts payable | \$4,990 [SEP]  [row 2]: Partners payable | \$1,117 [SEP]  [row 3]: Operating lease liabilities, current | \$1,367 [SEP]  [row 4]: Accrued expenses and other current liabilities | \$19,952 [SEP]  [row 5]: Total current liabilities | \$27,026 [SEP]  [row 6]: Operating lease liabilities, non-current | \$15,301 [SEP]  [row 7]: Long-term debt | \$9,923 [SEP]  [row 8]: Other liabilities | \$7,764 [SEP]  [row 9]: Total liabilities | \$60,014 [SEP]   [Time]: December 31, 2023 [SEP] [row 0]: Stockholders' equity [SEP]  [row 1]: Additional paid-in capital | \$73,253 [SEP]  [row 2]: Accumulated other comprehensive loss | (\$2,155) [SEP]  [row 3]: Retained earnings | \$82,070 [SEP]  [row 4]: Total stockholders' equity | \$153,168 [SEP]   [Time]: December 31, 2022 [SEP] [row 0]: Stockholders' equity [SEP]  [row 1]: Additional paid-in capital | \$64,444 [SEP]  [row 2]: Accumulated other comprehensive loss | (\$3,530) [SEP]  [row 3]: Retained earnings | \$64,799 [SEP]  [row 4]: Total stockholders' equity | \$125,713 [SEP]   [Time]: December 31, 2023 [SEP] [row 0]: Total liabilities and stockholders' equity | \$229,623 [SEP]   [Time]: December 31, 2022 [SEP] [row 0]: Total liabilities and stockholders' equity | \$185,727 [SEP]
    \end{quote}
\end{GenerateSyntheticTransactionData1}

\newpage
\begin{GenerateSyntheticTransactionData2}

Your Output: 
Trasactons For 2023:
[contributing to row 1]: Cash and cash equivalents | \$41,862 [SEP]
1. Advertising revenue of \$22,000 was collected from clients, with direct cash payment by the clients.
2.The sale of short-term investments generated \$22,862 in cash income.
3.Employee wages resulted in a cash outflow of \$3,000.
[Explanation: 41,862 = 22,000 + 22,862 - 3,000]

[contributing to row 2]: Marketable securities
4.Purchase of short-term government securities \$23,541.
[Explanation: \$23,541=\$23,541]

[contributing to row 3]: Accounts receivable, net 
5.Advertisers were invoiced for Q4 campaigns, totaling \$18,000.
6.A write-off of uncollectible accounts was recorded, totaling \$1,831.
[Explanation: \$16,169 = \$18,000- \$1,831]

[contributing to row 4]: Prepaid expenses and other current assets 
7.\$1,000 prepayment for upcoming marketing events.
8.Advance payment for annual cloud service subscriptions for \$2,793.
[Explanation:\$3,793= \$1,000 + \$2,793]

[contributing to row 5]: Total current assets
[Explanation: \$85,365 = 41,862 + \$23,541 + \$16,169 + \$3,793]

[contributing to row 6]: Non-marketable equity securities  
9.There was no major change except for minor revaluations or adjustments, resulting in a net increase of \$6,141.
[Explanation:\$6,141=\$6,141]

[contributing to row 7]: Property and equipment, net 
10.Capital expenditure on new data centers and office equipment totaled \$95,000.
11.Brought property this year for \$1,587.
[Explanation:\$96,587 = \$95,000+\$1,587]

[contributing to row 8]:Operating lease right-of-use assets  
12.Recognized additional lease assets worth \$14,000 from new agreements.
13.Amortization of existing lease assets reduced the value by \$706.
[Explanation: \$13,294 = \$14,000 - \$706]

[contributing to row 9]: Intangible assets, net  
14.Development costs that meet the capitalization criteria are included in intangible assets \$788.
[Explanation: \$788 = \$788]

[contributing to row 10]:Goodwill 
15.Acquisition of a small tech firm increased goodwill to \$20,654.
[Explanation: \$20,654 = \$20,654]

[contributing to row 11]:Other assets 
16.Other assets in the company totaled \$6,794.
[Explanation: \$6,794 = \$6,794]

[contributing to row 12]:Total assets 
[Explanation: \$229,623 =\$85,365 + \$6,141 + \$96,587 + \$13,294 + \$788+ \$20,654 + \$6,794]

[contributing to row 1]:Accounts payable 
17.Payments to vendors reduced accounts payable by \$151.
18.New purchases on credit added \$5,000 to accounts payable.
[Explanation: \$4,849 = \$5,000-\$151]

[contributing to row 2]Partners payable
19.Partners payable for this year totaled \$863.
[Explanation: \$863 = \$863]

[contributing to row 3] Operating lease liabilities, current  
20.Payments toward lease obligations totaled \$1,623.
[Explanation: \$1,623 = \$1,623]

[contributing to row 4]: Accrued expenses and other current liabilities  
21.Additional accrued expenses for Q4 salaries and benefits amounted to \$25,000.
22.Payment of \$375 toward previously accrued liabilities.
[Explanation: \$24,625 = \$25,000 - \$375]

[contributing to row 5]: Total current liabilities 
[Explanation: \$31,960= \$4,849 + \$863 + \$1,623 + \$24,625]

[contributing to row 6]: Operating lease liabilities, non-current  
23.New lease agreements added \$18,000.
24.Amortization of long-term liabilities reduced by \$774.
[Explanation: \$17,226 = \$18,000-\$774]

[contributing to row 7]: Long-term debt 
25.Issued new bonds worth \$18,385 for business expansion.
[Explanation: \$18,385 = \$18,385]

[contributing to row 8]: Other liabilities 
26.Other liabilities for the company totaled 8,884.
[Explanation: \$8,884 = \$8,884]

[row 9]: Total liabilities | \$76,455 
[Explanation: \$76,455 = \$31,960 + \$17,226 + \$18,385 + \$8,884]

[contributing to row 1]: Additional paid-in capital 
27.Stock issuance generated \$73,253 in additional paid-in capital.
[Explanation: \$73,253 = \$73,253]

[contributing to row 2]: Accumulated other comprehensive loss 
28.Decrease due to foreign currency adjustments and revaluation of securities amounted to \$2,155.
[Explanation: (\$2,155) = (\$2,155)]

[contributing to row 3]: Retained earnings 
29.Retained Earning for the company totaled 82,070.
[Explanation: \$82,070 = 82,070]

[contributing to row 4]: Total stockholders' equity 
[Explanation: \$153,168 = \$73,253 + (\$2,155) + \$82,070]

[contributing to row 0]: Total liabilities and stockholders' equity | \$229,623
[Explanation: \$229,623 = \$76,455 + \$153,168]

Now let's begin! Output your transactions which aligns with the table !

Input: \{ \}

\end{GenerateSyntheticTransactionData2}

\begin{TemplatesforErrorInjectioninFinancialStatements1}
    We inject errors into textual financial statements and generate corresponding judgments and explanations for incorrect entries. The template is constructed as follows:

\begin{quote}
You will help generate modified financial statements with specific types of errors for financial auditing purposes. Based on the input data from the original financial statement, you need to create a version with deliberate errors, as well as identify and explain the errors and provide corrections. The error types are provided as below: <1> Missing Row: Delete a row containing a specific account with values (e.g., an account under "Operating Activities," "Investing Activities," or "Financing Activities" in the Consolidated Statements of Cash Flows) but avoid removing key totals like "Total Revenue," "Total Expenses," "Total Assets," "Total Liabilities," "Total Equity," or "Net Cash Provided by Operating/Investing/Financing Activities.". <2> Numerical Error: Modify the value of a specific account (e.g., "Accounts Receivable" under Assets) by changing it to an incorrect amount. <3> Redundant Row: Add a new row with a logical account name and amount that fits the respective category (e.g., adding "Software Licenses" under Assets or "Deferred Revenue" under Liabilities). <4> Misclassification: Move a row from one category to another where it logically does not belong (e.g., relocating "Accounts Payable" from Liabilities to Assets in the Balance Sheet, or shifting an expense item under Revenue in the Consolidated Statement of Income). The outputs should consist of the following components:

"Modified Financial Statement with Errors": Introduce one of the following errors into the financial statement and output the changed statement as a table similar to input. You can use any of the type of errors introduced above by deleting rows, change a number, add an extra row, or move a row to a wrong place.

"General Judgment": Always output 'Incorrect' as you are introducing errors to the right table.

"Error Identification": Output the error type you have introduced together with the problematic entry row index. Generate your results with "Error type"

"Error Resolution": Provide a detailed explanation of the specific error(s), including why it is incorrect. For example: "The 'Inventory' item was wrong; it should be \$2000, and the calculation should be..", "The 'Accounts Liabilities' item was erroneously classified under Assets.", "Receivables (with the amount) was omitted." Then explain the error and suggest specific corrections to resolve the error(s). For example: "Accounts Liabilities does not belong under Assets; it should be moved to Liabilities." "Add 'Receivables' (with the amount) under Current Assets."

"Standards Citation": Cite the relevant Financial Accounting 
Standards Board (FASB) guidance related to the error, including specific references to sections, if applicable.

Your task is to generate outputs for each section described above in detail. Make sure the modifications are logical, align with accounting principles, and provide concise yet accurate error descriptions.

Here is an example output for your reference: 

Input: 
"Input Table": [Time]: September 30, 2023 [SEP] [row 0]: Current assets: [SEP]  [row 1]: Cash and cash equivalents | \$29,965 [SEP]  [row 2]: Marketable securities | \$31,590 [SEP]  [row 3]: Accounts receivable, net | \$29,508 [SEP]  [row 4]: Vendor non-trade receivables | \$31,477 [SEP]  [row 5]: Inventories | \$6,331 [SEP]  [row 6]: Other current assets | \$14,695 [SEP]  [row 7]: Total current assets | \$143,566 [SEP]  [row 8]: Non-current assets: [SEP]  [row 9]: Marketable securities | \$100,544 [SEP]  [row 10]: Property, plant and equipment, net | \$43,715 [SEP]  [row 11]: Other non-current assets | \$64,758 [SEP]  [row 12]: Total non-current assets | \$209,017 [SEP]  [row 13]: Total assets | \$352,583 [SEP]  [row 14]: Current liabilities: [SEP]  [row 15]: Accounts payable | \$62,611 [SEP]  [row 16]: Other current liabilities | \$58,829 [SEP]  [row 17]: Deferred revenue | \$8,061 [SEP]  [row 18]: Commercial paper | \$5,985 [SEP]  [row 19]: Term debt | \$9,822 [SEP]  [row 20]: Total current liabilities | \$145,308 [SEP]  [row 21]: Non-current liabilities: [SEP]  [row 22]: Term debt | \$95,281 [SEP]  [row 23]: Other non-current liabilities | \$49,848 [SEP]  [row 24]: Total non-current liabilities | \$145,129 [SEP]  [row 25]: Total liabilities | \$290,437 [SEP]  [row 26]: Commitments and contingencies | [SEP]  [row 27]: Shareholders' equity: [SEP]  [row 28]: Common stock and additional paid-in capital | \$73,812 [SEP]  [row 29]: Accumulated deficit | \$(214) [SEP]  [row 30]: Accumulated other comprehensive loss | \$(11,452) [SEP]  [row 31]: Total shareholders' equity | \$62,146 [SEP]  [row 32]: Total liabilities and shareholders' equity | \$352,583 [SEP]  
\end{quote}
\end{TemplatesforErrorInjectioninFinancialStatements1}

\begin{TemplatesforErrorInjectioninFinancialStatements2}
\begin{quote}

"Transaction data": Transactions for 2023 uff1a  [contributing to row 1]: Cash and cash equivalents | \$29,965 [SEP] 1. Apple launched a new iPhone model, generating revenue of \$18,000, with direct cash payment by the customers. 2. The sale of short-term investments resulted in a cash income of \$15,000. 3. Vendor payments and operational expenses resulted in a cash outflow of \$3,035. [Explanation: \$29,965 = \$18,000 + \$15,000 - \$3,035]  [contributing to row 2]: Marketable securities  4.The company purchased various securities totaling \$31,590. [Explanation: \$31,590=\$31,590]  [contributing to row 3]: Accounts receivable, net  5. Income from iPhone sales yet to be received, billed at \$32,500. 6. A write-off of uncollectible accounts was done, amounting to \$2,992. [Explanation: \$29,508 = \$32,500 - \$2,992]  [contributing to row 4]: Vendor non-trade receivables  7. Receivables from partners such as app developers on the App store, totaling \$31,477. [Explanation: \$31,477 = \$31,477]  [contributing to row 5]: Inventories  8. Production of new iPhone and Mac computers added \$6,331 to inventories. [Explanation: \$6,331= \$6,331]  [contributing to row 6]: Other current assets  9. Other current assets for the company, consisting of prepaid expenses and advances, totaling \$14,695. [Explanation: \$14,695 = \$14,695]  [contributing to row 7]: Total current assets [Explanation: \$143,566 = 29,965 + \$31,590 + \$29,508 +6,331 +14,695]  [contributing to row 9]: Marketable securities - non-current   10. The company made long-term investments totaling \$100,544. [Explanation: \$100,544 = \$100,544]  [contributing to row 10]: Property, plant and equipment, net   11. Capital expenditure on new retail stores and updating manufacturing equipment, totaling \$43,715. [Explanation: \$43,715 = \$43,715]  [contributing to row 11]: Other non-current assets   12. Other non-current assets for the company totaled \$64,758. [Explanation: \$64,758 = \$64,758]  [contributing to row 12]: Total non-current assets  [Explanation: \$209,017 = \$100,544 + \$43,715 + \$64,758]  [contributing to row 13]: Total assets  [Explanation: \$352,583 = \$143,566 + \$209,017]  [contributing to row 15]: Accounts payable  13. Payments to vendors reduced accounts payable by \$12,000. 14. New purchases on credit added \$74,611 to accounts payable. [Explanation: \$62,611 = \$74,611 - \$12,000]  [contributing to row 16]: Other current liabilities  15. Accrued expenses and other outstanding payments for the company amounted to \$58,829. [Explanation: \$58,829 = \$58,829]  [contributing to row 17]: Deferred revenue   16. Revenue deferred for services to be provided in the future totaled \$8,061.  [contributing to row 18]: Commercial paper  17. Short-term debt issued by the company, totaling \$5,985. [Explanation: \$5,985 = \$5,985]  [contributing to row 19]: Term debt - current   18. Current portion of long-term term debt, amounting to \$9,822. [Explanation: \$9,822 = \$9,822]  [contributing to row 20]: Total current liabilities  [Explanation: \$145,308= \$62,611+ \$58,829 + \$8,061 +\$5,985 + \$9,822]  [contributing to row 22]: Term debt - non-current   19. The company issued long-term bonds worth \$95,281 for business expansion. [Explanation: \$95,281 = \$95,281]  [contributing to row 23]: Other non-current liabilities   20. Other long-term liabilities, including pension obligations, amounted to \$49,848. [Explanation: \$49,848 = \$49,848]  [contributing to row 24]: Total non-current liabilities  [Explanation: \$145,129 = \$95,281 + \$49,848]  [contributing to row 25]: Total liabilities  [Explanation: \$290,437 = \$145,308 + \$145,129]  [contributing to row 28]: Common stock and additional paid-in capital  21. Stock issuance generated an additional \$73,812 in paid-in capital. [Explanation: \$73,812 = \$73,812]  [contributing to row 29]: Accumulated deficit   22. There was a net loss for the company during this period, amounting to \$214. [Explanation: (\$214) = (\$214)]  [contributing to row 30]: Accumulated other comprehensive loss  23. This reflects an increase of loss from various sources such as foreign currency translation, totaling (\$11,452) [Explanation: (\$11,452) = (\$11,452)]  [contributing to row 31]: Total shareholders' equity  [Explanation: \$62,146 = \$73,812 - \$214 - \$11,452]  [contributing to row 32]: Total liabilities and shareholders' equity  [Explanation: \$352,583 = \$290,437 + \$62,146]

\end{quote}
\end{TemplatesforErrorInjectioninFinancialStatements2}

\begin{TemplatesforErrorInjectioninFinancialStatements3}
You Ideal output in the json format: 
{{
    "Modified Financial Statement with Errors": "[Time]: September 30, 2023 [SEP] [row 0]: Current assets: [SEP]  [row 1]: Cash and cash equivalents | \$29,965 [SEP]  [row 2]: Marketable securities | \$31,590 [SEP]  [row 3]: Vendor non-trade receivables | \$31,477 [SEP]  [row 4]: Inventories | \$6,331 [SEP]  [row 5]: Other current assets | \$14,695 [SEP]  [row 6]: Total current assets | \$114,058 [SEP]  [row 7]: Non-current assets: [SEP]  [row 8]: Accounts receivable, net | \$29,508 [SEP] [row 9]: Marketable securities | \$100,544 [SEP]  [row 10]: Property, plant and equipment, net | \$43,715 [SEP]  [row 11]: Other non-current assets | \$64,758 [SEP]  [row 12]: Total non-current assets | \$238,525 [SEP]  [row 13]: Total assets | \$352,583 [SEP]  [row 14]: Current liabilities: [SEP]  [row 15]: Accounts payable | \$62,611 [SEP]  [row 16]: Other current liabilities | \$58,829 [SEP]  [row 17]: Deferred revenue | \$8,061 [SEP]  [row 18]: Commercial paper | \$5,985 [SEP]  [row 19]: Term debt | \$9,822 [SEP]  [row 20]: Total current liabilities | \$145,308 [SEP]  [row 21]: Non-current liabilities: [SEP]  [row 22]: Term debt | \$95,281 [SEP]  [row 23]: Other non-current liabilities | \$49,848 [SEP]  [row 24]: Total non-current liabilities | \$145,129 [SEP]  [row 25]: Total liabilities | \$290,437 [SEP]  [row 26]: Commitments and contingencies | [SEP]  [row 27]: Shareholders' equity: [SEP]  [row 28]: Common stock and additional paid-in capital | \$73,812 [SEP]  [row 29]: Accumulated deficit | \$(214) [SEP]  [row 30]: Accumulated other comprehensive loss | \$(11,452) [SEP]  [row 31]: Total shareholders' equity | \$62,146 [SEP]  [row 32]: Total liabilities and shareholders' equity | \$352,583 [SEP]",
    "General Judgement": "Incorrect",
    "Error Identification": {{"Error Type": Misclassification, "Problematic Entry": Row 8}},
    "Error Resolution": "The Financial Statement mistakenly wrote accounts receivable, net into non-current assets. Move accounts receivable, net to the current assets item, delete it from Non-current assets, and then recalculate the total current assets and total non-current assets amounts.",
    "Standards Citation": "Current assets generally include: Receivables from officers, employees, affiliates, and others, if collectible in the ordinary course of business within a year."
}}

Now, let's begin! Remember to follow the output format that is provided by the few-shot example. In this round, you need to generate the error of \{Current error\}

Input: 
"Input Table": \{Current Table\}
        
"Transaction data": \{Current Transaction\}
\end{TemplatesforErrorInjectioninFinancialStatements3}

\begin{TemplatesforPromptingSotaLLMstoServeasAuditors1}

We use the following prompt to evaluate the LLMs' ability to audit financial statements.

\begin{quote}
    You will serve as a financial statement auditor helping to identify, explain, and correct intentional errors introduced into financial statements based on provided data. Your input will be an incorrect financial statement with corresponding transactions. The errors introduced into financial statements belong to one of the following types: <1> Missing Row: Delete a row containing a specific account with values (e.g., an account under "Operating Activities," "Investing Activities," or "Financing Activities" in the Consolidated Statements of Cash Flows) but avoid removing key totals like "Total Revenue," "Total Expenses," "Total Assets," "Total Liabilities," "Total Equity," or "Net Cash Provided by Operating/Investing/Financing Activities.". <2> Numerical Error: Modify the value of a specific account (e.g., "Accounts Receivable" under Assets) by changing it to an incorrect amount. <3> Redundant Row: Add a new row with a logical account name and amount that fits the respective category (e.g., adding "Software Licenses" under Assets or "Deferred Revenue" under Liabilities). <4> Misclassification: Move a row from one category to another where it logically does not belong (e.g., relocating "Accounts Payable" from Liabilities to Assets in the Balance Sheet, or shifting an expense item under Revenue in the Consolidated Statement of Income).

For each pair of input, you need to output your auditing results in the following format:

"General Judgment": Judge whether the table is problematic, output 'correct' if you believe that there is no errors, output 'incorrect' if you believe that there is errors. If you output 'correct', there is no need to output anything else. If you output 'incorrect', then you need to also output the following for each error you have identified. You should always use "Information for error 1" to wrap up your first identified error.

"Error Identification": Output the error type you have identified together with the problematic entry row index, for the row index, you should output in the form of 'Row N'.

"Error Resolution": Provide a detailed explanation of the specific error(s), including why it is incorrect. 

"Standards Citation": Cite the relevant Financial Accounting 
Standards Board (FASB) guidance related to the error, including specific references to sections, if applicable.

"Corrected Statements": Provide the corrected financial statement after addressing the identified errors.

Your task is to generate outputs for each section described above in detail. Make sure the modifications are logical, align with accounting principles, and provide concise yet accurate error descriptions.

Here is an example output for your reference: 

Input:
"Input Table": "[Time]: September 30, 2023 [SEP] [row 0]: Current assets: [SEP]  [row 1]: Cash and cash equivalents | \$29,965 [SEP]  [row 2]: Marketable securities | \$31,590 [SEP]  [row 3]: Vendor non-trade receivables | \$31,477 [SEP]  [row 4]: Inventories | \$6,331 [SEP]  [row 5]: Other current assets | \$14,695 [SEP]  [row 6]: Total current assets | \$114,058 [SEP]  [row 7]: Non-current assets: [SEP]  [row 8]: Accounts receivable, net | \$29,508 [SEP] [row 9]: Marketable securities | \$100,544 [SEP]  [row 10]: Property, plant and equipment, net | \$43,715 [SEP]  [row 11]: Other non-current assets | \$64,758 [SEP]  [row 12]: Total non-current assets | \$238,525 [SEP]  [row 13]: Total assets | \$352,583 [SEP]  [row 14]: Current liabilities: [SEP]  [row 15]: Accounts payable | \$62,611 [SEP]  [row 16]: Other current liabilities | \$58,829 [SEP]  [row 17]: Deferred revenue | \$8,061 [SEP]  [row 18]: Commercial paper | \$5,985 [SEP]  [row 19]: Term debt | \$9,822 [SEP]  [row 20]: Total current liabilities | \$145,308 [SEP]  [row 21]: Non-current liabilities: [SEP]  [row 22]: Term debt | \$95,281 [SEP]  [row 23]: Other non-current liabilities | \$49,848 [SEP]  [row 24]: Total non-current liabilities | \$145,129 [SEP]  [row 25]: Total liabilities | \$290,437 [SEP]  [row 26]: Commitments and contingencies | [SEP]  [row 27]: Shareholders' equity: [SEP]  [row 28]: Common stock and additional paid-in capital | \$73,812 [SEP]  [row 29]: Accumulated deficit | \$(214) [SEP]  [row 30]: Accumulated other comprehensive loss | \$(11,452) [SEP]  [row 31]: Total shareholders' equity | \$62,146 [SEP]  [row 32]: Total liabilities and shareholders' equity | \$352,583 [SEP]",

"Transaction data": Transactions for 2023: 
1. Apple launched a new iPhone model, generating revenue of \$18,000, with direct cash payment by the customers.
2. The sale of short-term investments resulted in a cash income of \$15,000.
3. Vendor payments and operational expenses resulted in a cash outflow of \$3,035.
4.The company purchased various securities totaling \$31,590
5. Income from iPhone sales yet to be received, billed at \$32,500.
6. A write-off of uncollectible accounts was done, amounting to \$2,992.
7. Receivables from partners such as app developers on the App store, totaling \$31,477.
8. Production of new iPhone and Mac computers added \$6,331 to inventories.
9. Other current assets for the company, consisting of prepaid expenses and advances, totaling \$14,695
10. The company made long-term investments totaling \$100,544.
11. Capital expenditure on new retail stores and updating manufacturing equipment, totaling \$43,715.
12. Other non-current assets for the company totaled \$64,758.
13. Payments to vendors reduced accounts payable by \$12,000.
14. New purchases on credit added \$74,611 to accounts payable.
15. Accrued expenses and other outstanding payments for the company amounted to \$58,829.
16. Revenue deferred for services to be provided in the future totaled \$8,061.
17. Short-term debt issued by the company, totaling \$5,985.
18. Current portion of long-term term debt, amounting to \$9,822.
19. The company issued long-term bonds worth \$95,281 for business expansion.
20. Other long-term liabilities, including pension obligations, amounted to \$49,848.
21. Stock issuance generated an additional \$73,812 in paid-in capital.
22. There was a net loss for the company during this period, amounting to \$214.
23. This reflects an increase of loss from various sources such as foreign currency translation, totaling \$11,452
\end{quote}
\end{TemplatesforPromptingSotaLLMstoServeasAuditors1}

\begin{TemplatesforPromptingSotaLLMstoServeasAuditors2}
\begin{quote}
You Ideal output in the json format: 

{{
    "General Judgment": "Incorrect",
    "Information for error 1":{{
        "Error Identification": {{
            "Error Type": Misclassification, 
            "Problematic Entry": Row 8
        }},
        "Error Resolution": "The Financial Statement mistakenly wrote accounts receivable, net into non-current assets. Move accounts receivable, net to the current assets item, delete it from Non-current assets, and then recalculate the total current assets and total non-current assets amounts.",
        "Standards Citation": "Current assets generally include: Receivables from officers, employees, affiliates, and others, if collectible in the ordinary course of business within a year.",
        "Corrected Statements": "[Time]: September 30, 2023 [SEP] [row 0]: Current assets: [SEP]  [row 1]: Cash and cash equivalents | \$29,965 [SEP]  [row 2]: Marketable securities | \$31,590 [SEP]  [row 3]: Accounts receivable, net | \$29,508 [SEP]  [row 4]: Vendor non-trade receivables | \$31,477 [SEP]  [row 5]: Inventories | \$6,331 [SEP]  [row 6]: Other current assets | \$14,695 [SEP]  [row 7]: Total current assets | \$143,566 [SEP]  [row 8]: Non-current assets: [SEP]  [row 9]: Marketable securities | \$100,544 [SEP]  [row 10]: Property, plant and equipment, net | \$43,715 [SEP]  [row 11]: Other non-current assets | \$64,758 [SEP]  [row 12]: Total non-current assets | \$209,017 [SEP]  [row 13]: Total assets | \$352,583 [SEP]  [row 14]: Current liabilities: [SEP]  [row 15]: Accounts payable | \$62,611 [SEP]  [row 16]: Other current liabilities | \$58,829 [SEP]  [row 17]: Deferred revenue | \$8,061 [SEP]  [row 18]: Commercial paper | \$5,985 [SEP]  [row 19]: Term debt | \$9,822 [SEP]  [row 20]: Total current liabilities | \$145,308 [SEP]  [row 21]: Non-current liabilities: [SEP]  [row 22]: Term debt | \$95,281 [SEP]  [row 23]: Other non-current liabilities | \$49,848 [SEP]  [row 24]: Total non-current liabilities | \$145,129 [SEP]  [row 25]: Total liabilities | \$290,437 [SEP]  [row 26]: Commitments and contingencies | [SEP]  [row 27]: Shareholders' equity: [SEP]  [row 28]: Common stock and additional paid-in capital | \$73,812 [SEP]  [row 29]: Accumulated deficit | \$(214) [SEP]  [row 30]: Accumulated other comprehensive loss | \$(11,452) [SEP]  [row 31]: Total shareholders' equity | \$62,146 [SEP]  [row 32]: Total liabilities and shareholders' equity | \$352,583 [SEP]  "
    }}
    "Information for error 2":{{
        ......
    }}
}}

An example for correct tables: if your output is correct just output :
{{
    "General Judgement": "Correct",
}}

Now, let's begin! Remember to follow the output format directly as json format. Do not output anything else except the json output.

Input: 
"Input Table": \{ \}
        
"Transaction data": \{ \}
\end{quote}
\end{TemplatesforPromptingSotaLLMstoServeasAuditors2}

\end{document}